# Ground-based Gamma-ray Telescopes as Ground Stations in Deep-Space Lasercom

Alberto Carrasco-Casado
Space Communication Systems Laboratory
National Institute of Information and Communications Technology
4-2-1 Nukui-Kitamachi, Koganei, Tokyo 184-8795, Japan
alberto@nict.go.jp

José Manuel Sánchez-Pena, Ricardo Vergaz
Electronic Technology Department, Carlos III University of Madrid
Av. de la Universidad 30, Madrid 28911, Spain

*Abstract*— As the amount of information to be transmitted from deep-space rapidly increases, the radiofrequency technology has become a bottleneck in space communications. RF is already limiting the scientific outcome of deep-space missions and could be a significant obstacle in the developing of manned missions. Lasercom holds the promise to solve this problem, as it will considerably increase the data rate while decreasing the energy, mass and volume of onboard communication systems. A key strategy to optimize the lasercom technology is to shift the complexity from the space systems towards the ground systems. In RF deep-space communications, where the received power is the main limitation, the traditional approach to boost the data throughput has been increasing the receiver's aperture, e.g. the 70-m antennas in the NASA's Deep Space Network. Optical communications also can benefit from this strategy, thus 10-m class telescopes have typically been suggested to support future deep-space links. However, the cost of big telescopes increase exponentially with their aperture, and new ideas are needed to optimize this ratio. Here, the use of ground-based gamma-ray telescopes, known as Cherenkov telescopes, is suggested. These are optical telescopes designed to maximize the receiver's aperture at a minimum cost with some relaxed requirements. As they are used in an array configuration and multiple identical units need to be built, each element of the telescope is designed to minimize its cost. Furthermore, the native array configuration would facilitate the joint operation of Cherenkov and lasercom telescopes. These telescopes offer very big apertures, ranging from several meters to almost 30 meters, which could greatly improve the performance of optical ground stations. The key elements of these telescopes have been studied applied to lasercom, reaching the conclusion that it could be an interesting strategy to include them in the future development of an optical deep-space network.

*Keywords*— *Free-space optical communication; Deep-space communication; Cherenkov telescopes*

## I. Introduction

Scientific data to be transmitted from deep-space missions have been increasing since the beginning of the space age, and it will continue to grow in the future [1]. A communication bottleneck has already been reached, meaning that an important amount of meaningful data is lost because of communication limitations. With current radiofrequency (RF) technology, a spacecraft lifetime would allow to map below ten percent of the surface of Mars, whereas lasercom technology could do it at a rate of over a hundred times faster. Furthermore, future manned missions will require a qualitative leap in the way data is interchanged with the Earth.

Lasercom holds the promise to fill this gap and bring a new kind of high-bitrate communication, increasing the bandwidth and reducing the power, mass and size of spacecraft payloads. The narrow divergence of optical transmission suppose an even more apparent advantage in the vast distances of deep-space links. A reduction of over 4 orders of magnitude in the footprint of a beam reaching the Earth from Mars would be possible using optical frequencies instead of RF.

Large apertures have always been necessary for RF ground stations to receive the low-power signals received from deep-space. In Free-Space Optical Communication (FSOC), the same strategy can be used to overcome the big losses, and also to follow the rule of shifting the complexity from the Space to the Earth. In this paper, big-aperture IACTs (Imaging Atmospheric Cherenkov Telescopes) are proposed to be reutilized as optical ground terminals for deep-space FSOC.

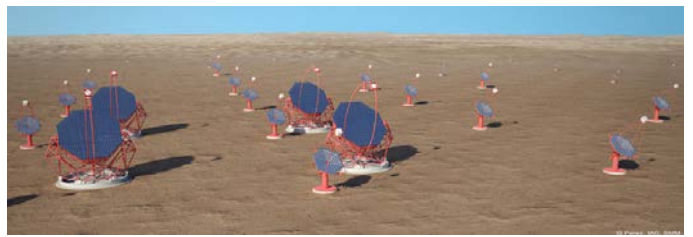

Fig. 1. Artistic illustration of the Cherenkov Telescope Array [2].

## II. CTA project

IACTs are optical telescopes design to study high-energy astronomy (especially gamma and cosmic rays) on the ground. These telescopes do not detect high-energy radiation directly, but only its effects after interacting with the upper atmosphere. Gamma and cosmic rays interact with atmospheric molecules and as a consequence produce Cherenkov radiation in the visible spectrum. These photons of light, with abundance of ultraviolet and blue components, are detected on the ground by IACTs or Cherenkov telescopes by means of their big segmented mirrors.

The Cherenkov Telescope Array (CTA) (fig. 1) is an international collaboration to build a big number of IACTs in an observatory in the northern hemisphere (in La Palma, Spain) and another one in the southern hemisphere (in Paranal grounds, Chile) (fig. 2). This 200 M€ project involves 31 countries and its big interest is based on the success of current Cherenkov telescopes operating in different observatories around the world.

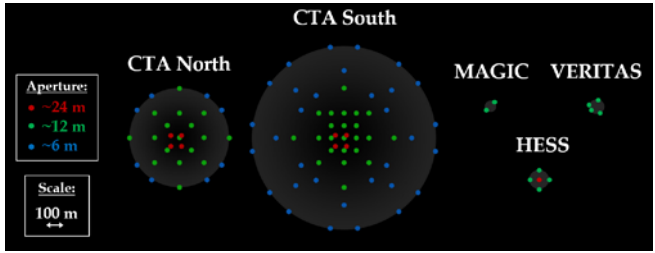

Fig. 2. CTA sites compared with current Cherenkov observatories.

In the northern site (located at the Spanish observatory), the Institute of Astrophysics of Canarias (IAC) signed this year a collaborative agreement with the University of Tokyo, as the leader of a large consortium of universities and research institutes which are members of the CTA Consortium, to carry out the construction of the first 24-m telescope within 2015-2016.

Cherenkov astronomy shares a common limitation with deep-space FSOC, i.e. the photon-starved detection regime. The solution followed in IACTs has been traditionally the utilization of very big apertures on the ground. In CTA, the same strategy will be followed, as well as the replication of elements in a big array of telescopes. This approach is also very convenient for deep-space FSOC, where 10-m class telescopes have been suggested in previous feasibility studies to sustain communication links from Mars and beyond.

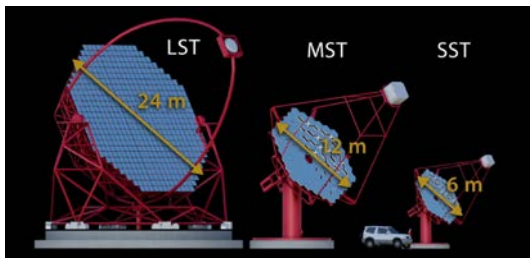

Fig. 3. Types of CTA telescopes [3].

Although there are several array alternatives under study for CTA, the southern site will likely consist of over 100 telescopes and over 20 telescopes in the northern one. Three different IACT sizes (fig. 3) are foreseen to be developed. In each site, there will be many dozens of SST (Small Size Telescopes) with a diameter of 6 m and separated by 70 m; several tens of MST (Medium Size Telescopes) with a diameter of 12 m and separated by 100 m; and around four units of LST (Large Size Telescopes) with a diameter of 24 m and separated by 100 m.

## III. MOTIVATIONS FOR THIS PROPOSAL

There are a number of motivations to justify the interest of this proposal. Below, a brief overview of the most important ones is presented.

- Deep-space lasercom is limited by the received power. An important improvement would come from the increase of the receiver aperture, and CTA telescopes offer bigger apertures than any previous design for FSOC.
- CTA telescopes work as an array topology in a native way. This could be used for arbitrarily increase the receiver's aperture by replicating telescopes. Thus, the main advantage (big apertures) is also scalable.
- A big increase in the receiver's aperture as the one proposed here could be used for alleviating the requirements in the spacecraft terminals, shifting the complexity towards the ground stations.
- The CTA deployment phase will be of such magnitude that it could be seen as a mass production of telescopes, where the unitary costs are greatly reduced. The costs of the infrastructure would be also lower for being shared.
- FSOC telescopes generally do not need the image-formation optical quality of conventional astronomical telescopes. IACTs are also designed with lower requirements, which greatly reduce their costs.
- Cherenkov astronomy shares with FSOC the same requirements regarding atmospheric conditions. CTA telescopes will be located in astronomical observatories with almost ideal atmospheric conditions.

## IV. IACT MIRRORS

The reflectance of IACT mirrors is a critical parameter in order to take advantage of the mass production of mirrors that will take place for CTA telescopes. Cherenkov telescopes are designed to receive visible radiation. However, FSOC uses near-infrared spectrum (mainly 1550 nm). Since this wavelength is far from Cherenkov radiation, IACTs mirrors are not characterized in this region. To assess the performance of Cherenkov telescopes when used for FSOC, several measurements were made over different mirrors of IACTs.

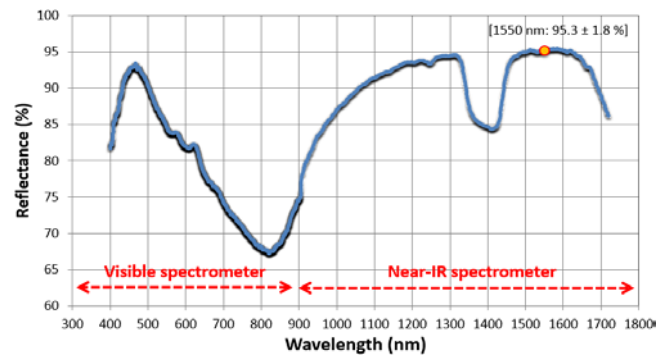

Fig. 4. Reflectance measured over MAGIC telescope mirrors.

There are two kinds of mirrors under study for CTA telescopes: metallic mirrors and glass mirrors. Firstly, MAGIC telescope mirrors reflectance were measured (fig. 4) as a good sample of diamond polishing technique for aluminum. MAGIC is the most advance IACT currently under operation. These measurements were carried out in La Palma observatory (Spain). Two different portable spectrometers were used for each spectral region: 200-1100 nm and 900-1700 nm. The former was used for verifying the measurements with previous ones (showing a very good agreement) and the latter for characterizing the mirror at the wavelength of interest. Although the MAGIC mirrors were not designed to optimize their behavior in the 1550 nm region, a reflectance of over 95 % was measured.

Other techniques are also under study for manufacturing CTA glass mirrors. Replica-based procedure is a good candidate due to its low cost when manufacturing spherical mirrors like the ones required for CTA. These techniques consists of conforming the mirror using a mold based on a honeycomb structure [3]. To avoid the degradation of the aluminum layer, and at the same time increasing the reflectance, interference dielectric layers are under study [4].

Several samples of the main candidate techniques for CTA were used to measure their reflectance at Cherenkov region as well as at 1550 nm (fig. 5). AR100 is based on an aluminum and a quartz layer (Al+SiO2), DH100 is made up by an aluminum layer with dielectric layer (SiO2+HfO2+SiO2), and DD040 is a dielectric mirror. AR100 and DH100 showed a reflectance over 90 % in FSOC wavelength, whereas DD040 did not reach 10 %, showing an excellent performance in the Cherenkov region, where it was optimized to operate. However, these kinds of mirrors can be tuned to optimize the reflectance at 1550 nm without any cost increase [5]. It can be concluded that all of the techniques under study for CTA can be utilized for fabricating mirrors for FSOC operation.

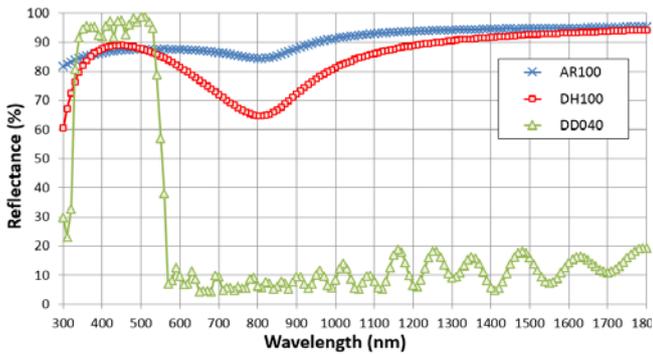

Fig. 5. Reflectance measured over several CTA mirror samples.

## V. Focusing an IACT

Another important difference between Cherenkov telescopes and FSOC telescopes is the distance to the object plane. FSOC telescopes, as well as astronomical telescopes, are designed to observe very distant objects. These objects can be considered to be at infinity, thus the photodetector is placed at the focal length of the telescope. Conversely, IACTs observe the atmosphere at a height of ~10 km, since the Cherenkov radiation is produced at a random height between 6 km and 20 km [6]. As they are only focused to observe at a given distance, any other distance will be defocused. This is the main cause of the relaxed optical requirements of IACTs.

As a consequence of the different distance to the object plane, placing the camera/detector in a different image plane for each application is required. This means that when adapting a Cherenkov telescope for FSOC operation, a displacement ε in the image plane, where the photodetector is placed instead of the original Cherenkov camera, will be needed. Using the image formation equation [7], this displacement can be calculated.

TABLE I. IMAGE PLANE DISPLACEMENT ε FOR FSOC ADAPTATION.

| Telescope | Diameter D (m) | f/D | Focal length f (m) | ε (cm) |
|---|---|---|---|---|
| MAGIC | 17 | 1 | 17 | 2.9 |
| CTA-SST | 6 | 0.5 | 12 | 1.4 |
| CTA-MST | 12 | 1 / 0.75 | 16 | 2.5 |
| CTA-LST | 24 | 1.25 | 30 | 9 |

Table I shows the displacement ε applied to MAGIC and CTA telescopes, with values ranging from 1.4 cm (SST) to 9 cm (LST). MAGIC-II telescope allows to shift its camera 30 cm for focusing and maintenance [8], being a normal feature in this kind of telescopes and allowing the FSOC adaptation. It is also important to stress that the space requirements to place the receiver's optical system will be in general much less strict when compared with an IACT. The reason is that Cherenkov telescopes use very big and heavy (several tons of weight) multi-pixel (several thousands of photomultipliers) cameras.

## VI. Field of View in IACTs

The Field of View (FoV) of a telescope shows the angular spread of an object when projected in the image plane. The optical resolution of the telescope, characterized by the Point Spread Function (PSF), determines the spatial distribution of the radiation received from a point source at infinity. Both concepts are very important when using a telescope for communications and are related through the background noise.

IACTs operate only at night. Conversely, FSOC telescopes also need to be operated during the day. Thus, the sunlight, reaching the telescope due to the scattering in the atmosphere, is the main source of background noise. This background light is detected by the receiver as a noise power $N_S$. According to eq. (1), $N_S$ depends on the sky radiance $L(\lambda,\theta,\varphi)$, the aperture diameter $D_r$, the FoV angle $\theta_{FOV}$, and the spectral bandwidth of the signal $\Delta\lambda$.

$$N_S = L(\lambda,\theta,\varphi)\cdot(\pi/4\cdot D_r\cdot \theta_{FOV})^2\cdot \Delta\lambda \qquad (1)$$

The FoV $\theta_{FOV}$ is proportional to the detector size and inversely proportional to the focal length. The detector size is the parameter that allows more control over the background noise in this kind of telescopes. This size should be as small as possible and it should match the PSF in order to minimize the received background noise and avoid to lose any signal photon.

To evaluate the optical performance of Cherenkov telescopes, two of them were simulated using ray-tracing software. Firstly, MAGIC-II (fig. 6) was modelled using OpticsLab from Science Lab Software, and also CTA-LST

(fig. 7) using OSLO (Optics Software for Layout and Optimization) from Lambda Research. The estimated PSF for both telescopes are 2.61 cm and 6.2 cm, respectively. Analyzing the results of these simulations, it could be concluded that IACTs are limited in their optical resolution by their geometrical aberrations, much more than the atmospheric turbulence or any diffraction effect.

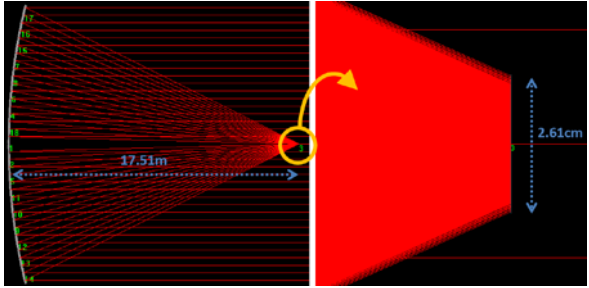

Fig. 6. Simulation of the optical resolution of MAGIC-II telescope.

Several different kinds of telescopes will be built in CTA. LST will have a parabolic profile (although spherical mirrors) and MST and SST will have two types of profiles: Davies-Cotton (DC) (based also in spherical mirrors and a primary focus design) and Schwarzschild-Couder (SC) (based in aspherical mirrors and a secondary focus design). SC type will also have two variations: British-French GATE (SST-SC GATE) [9], and Italian ASTRI (SST-SC ASTRI) [10]. After studying their design constraints, the SC type was chosen as the best candidate to be used for FSOC because of their better optical resolution, allowing up to one order of magnitude narrower FoV, which minimizes the background noise.

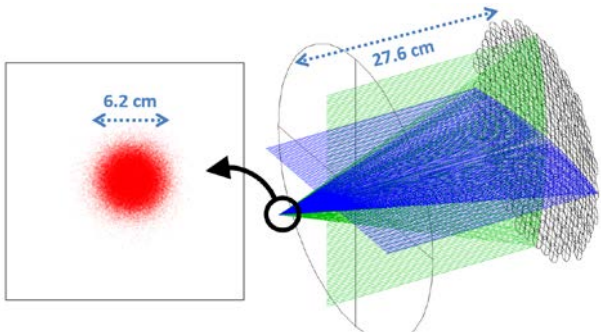

Fig. 7. Simulation of the optical resolution of CTA-LST telescope.

## VII. BACKGROUND NOISE

The background noise received by the CTA telescopes was estimated using eq. (1) in order to characterize their communications performance in terms of the Signal-to-Noise Ratio (SNR) in section VIII. The sky radiance $L(\lambda,\theta,\varphi)$ was calculated using MODTRAN (MODerate resolution atmospheric TRANsmission) software from Spectral Sciences. In this calculation, a minimum Sun-Earth-probe angle of 5° was used, and a maximum observation zenith angle of 70°. The conditions of La Palma observatory (Spain) were simulated, since it will be the site chosen for CTA-north.

The fig. 8 shows the sky radiance for a solar zenith angle of 45°, azimuthal angle of 0° and observation zenith angles of 0°-90° for a wavelength of 1550 nm. When limiting the observation zenith angle to 70° and Sun-Earth-probe angle to 5°, a maximum sky radiance of 430 µW/(cm$^2$·srad·µm) was obtained. This value was used as a worst case during daylight operation in eq. (1), getting a background noise $N_S$ of 0.183 nW for SST-SC ASTRI, 0.052 nW for SST-SC GATE, and 0.041 nW for MST-SC.

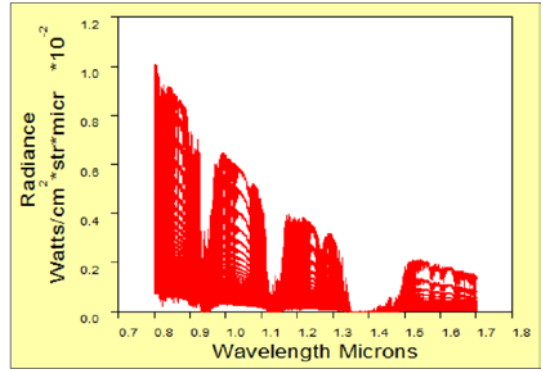

Fig. 8. Sky radiance for La Palma observatory (Spain).

## VIII. LINK BUDGET

A link budget has been simulated to obtain the SNR using the background noise calculation from section VII. Different scenarios were simulated as suggested by the OLSG (Optical Link Study Group) [11]: LEO, Lagrange points 1 and 2, Moon and Mars in conjunction and opposition. For each scenario, the SNR was obtained using each telescope SST-SC ASTRI, SST-SC GATE and MST-SC. Other link parameters different from the ground stations ones were taken from the OLSG report. For scenarios involving daytime, the worst-case background noise calculated in section VII was used. For nighttime scenarios, also a worst case was considered, assuming that Mars albedo enters the receiver's FoV.

TABLE II. SNR (dB) FOR EACH TELESCOPE AND SCENARIO.

|  | LEO | Moon | L1 | L2 | Mars opp. | Mars conj. |
|---|---|---|---|---|---|---|
| **SST-SC ASTRI** | 67.35 ✓ | 25.37 ✓ | 18.35 ✓ | 25.30 ✓ | 6.15 ✓ | -13.78 ✗ (48×) 3 ✓ |
| **SST-SC GATE** | 71.75 ✓ | 29.80 ✓ | 22.77 ✓ | 25.30 ✓ | 6.15 ✓ | -9.35 ✗ (18×) 3.2 ✓ |
| **MST-SC** | 80.07 ✓ | 50.03 ✓ | 31.09 ✓ | 25.30 ✓ | 6.15 ✓ | -1.04 ✗ (3×) 3.7 ✓ |

The table II shows a summary of the SNR for each scenario and each CTA selected telescope when adapted as a FSOC receiver. Between brackets, the number of telescopes in an array configuration is indicated, in order to close each link with a SNR > 3 dB when a single telescope is not enough. Only Mars-in-conjunction links (the worst-case scenario regarding free-space path loss and daylight background noise) are not fulfilled using one single telescope (an array of three elements of MST-SC would be the best solution). The rest of the scenarios are feasible with any CTA telescope.

It should be stressed that even the best CTA telescope regarding FoV (MST-SC, with a FoV of 0.3 mrad) is still around one order of magnitude worse than the performance of a FSOC telescope. Therefore, in order to make the most of CTA telescopes and minimize the background noise, further research on improvements of the aberration behavior should be carried out. It is also important to note that the cost of a MST telescope, as foreseen by CTA consortium [12], is two orders of magnitude lower than a FSOC ground station for Mars [11], even having a bigger aperture area.

## IX. CONCLUSIONS

A huge number of big-aperture segmented optical telescopes will be built under the CTA project in the next few years, with important unitary costs reduction when compared to ad-hoc developments. There are a number of characteristics that IACT and FSOC telescopes share, which suggests that adapting and reutilizing some Cherenkov telescopes as ground stations for deep-space lasercom would be an interesting idea to study. Here, some of the motivations to sustain this proposal were presented. Also, the two most important differences between IACT and FSOC telescopes were analyzed (namely, the working wavelength and the focusing goals) and it was concluded that they do not suppose an impediment when adapting these telescopes for communication purposes. IACT main limitation (the optical resolution due to the geometric aberration of spherical optics) was analyzed and worst-case simulations were carried out in order to assess the performance of IACTs as communication receivers. It was concluded that they could sustain almost all the simulated links, although there is still margin for improvement, especially regarding their aberration behavior, which limits the minimum FOV, and therefore the background noise coupled to the detector.